\documentclass{article}
\usepackage{spconf,amsmath,graphicx}

\usepackage{xcolor}
\usepackage{afterpage}
\usepackage{tabularx}
\usepackage{booktabs}
\usepackage{amsmath}
\usepackage{colortbl}

\usepackage{comment}
\usepackage{url}

\usepackage{makecell}

\makeatletter\renewcommand\paragraph{\@startsection{paragraph}{4}{\z@}
  {.5em \@plus1ex \@minus.2ex}{-.5em}{\normalfont\normalsize\bfseries}}\makeatother

\definecolor{palepurple}{RGB}{220, 198, 224}
\definecolor{paleyellow}{RGB}{255, 255, 153}
\definecolor{orange}{RGB}{255, 153, 0}
\definecolor{purple}{rgb}{0.5, 0.0, 0.5}

\newcommand\todo[1]{\textcolor{red}{#1}}

\definecolor{applegreen}{rgb}{0.55, 0.71, 0.0}
\definecolor{babyblueeyes}{rgb}{0.33, 0.49, 0.65}
\definecolor{candypink}{rgb}{0.89, 0.44, 0.48}

\DeclareMathOperator*{\argmax}{arg\,max}

\definecolor{pink}{rgb}{1.0, 0.75, 0.8}
\definecolor{piggypink}{rgb}{0.99, 0.87, 0.9}
\usepackage{soul} 
\sethlcolor{pink}

\title{Prompt Performance Prediction for Image Generation}

\name{
    \parbox{\linewidth}{\centering
    Nicolas BIZZOZZERO \qquad
    Ihab BENDIDI$^{1}$ \qquad 
    Olivier RISSER-MAROIX$^{2\dag}$ 
    }
   \thanks{
   $\dag$ Corresponding author: \texttt{orissermaroix@gmail.com}}
}

\address{
    \vspace{1mm}$^1$Minos Biosciences, Ecole Normale Supérieure, Paris, France\\
    \vspace{0.5mm}$^2$Artinity, Paris, France
}

\begin{document}
%
\maketitle
\begin{abstract}
The ability to predict the performance of a query before results are returned has been a longstanding challenge in Information Retrieval (IR) systems. 
Inspired by this task, we introduce, in this paper, a novel task called "Prompt Performance Prediction" that aims to predict the performance of a prompt, 
before obtaining the actual generated images. 
We demonstrate the plausibility of our task by measuring the correlation coefficient between predicted and actual performance scores across: three datasets containing pairs of prompts and generated images and three art domain datasets of real images and real user appreciation ratings. 
Our results show promising performance prediction capabilities, suggesting potential applications for optimizing user prompts. 
\end{abstract}

\section{Introduction}

In recent years, there have been remarkable advancements in image generation techniques, particularly those that are driven by textual inputs, commonly referred to as prompts. These innovations have been greatly facilitated by the use of pre-trained vision-language models. Notable among these models are CLIP \cite{radford2021learning} and ALIGN \cite{jia2021scaling}, as well as their derivatives \cite{goel2022cyclip, liang2022mind}. An important feature of image generation is its capacity to create entirely new content 
which contrasts with classical content-based image retrieval (CBIR) systems, which rely on existing images to provide user satisfaction.

Despite these advancements, a critical aspect remains less explored – predicting the effectiveness of prompts in generating a relevant image. 
This task, close to Query Performance Prediction (QPP) in traditional information retrieval (IR) \cite{carmel2012query, chen2023unified}, remains pertinent, prompting us to extend its scope into the image generation domain.  
To address this, we introduce a novel task, \textit{"Prompt Performance Prediction"} (PPP), focusing on the prediction of a prompt's performance before obtaining the generated outputs. 

In this paper, we propose a proactive approach to predicting prompt performance for image generation. 
We propose to use machine learning techniques to gauge the effectiveness of a prompt based on the relevance of generated images. 
The relevance is measured by various metrics such as aesthetic score and memorability, across three popular generative frameworks (DALL-E 2, Midjourney, and Stable Diffusion). Our extensive experiments establish a significant correlation between the predicted prompt performance scores and the actual performance observed, signifying the potency of our approach in accurately predicting prompt performance.

The implications of our research are manifold. By enabling users to gauge the potential effectiveness of their prompts, we provide a platform for improved content creation, advertising strategies, and user experience design. Furthermore, the PPP task provides valuable feedback for enhancing generative models themselves.

In this paper, we make the following contributions to the field of generative IR and query performance prediction:
\begin{itemize}
    \item Inspired by traditional Query Performance Prediction (QPP), we introduce the novel task of Prompt Performance Prediction (PPP). 
    \item To evaluate the feasibility of our task, we performed extensive baseline experiments on three distinct prompt-image datasets derived from popular generative systems and a complementary analysis on art domain.
\end{itemize}
The outcomes of these experiments validate the interest of this task while leaving ample room for future work: optimization of the PPP pipeline, integration in automatic prompt (query) reformulation, etc.

\section{Related Work}

Query Performance Prediction (QPP) has been extensively studied in the field of Information Retrieval (IR) as a means of estimating search effectiveness without relying on relevance information \cite{carmel2012query, chen2023unified}. 
The prediction can be performed either pre-retrieval or post-retrieval, each with its own distinct approaches and methodologies. 

In the context of image retrieval, recent work has introduced benchmark datasets, such as iQPP, to evaluate Image Query Performance Prediction methods \cite{chen2023unified}. However, these benchmarks are designed for traditional image retrieval systems and do not directly apply to generative image retrieval.

In computer vision, various approaches based on image features have been used to predict aesthetics \cite{murray2012ava, kong2016photo, schwarz2018will,  hentschel2022clip}, 
memorability \cite{isola2011understanding, fajtl2018amnet, ResMem2021, hagen2023image}, 
composition quality \cite{zhang2021image}, interestingness \cite{Gygli2013, constantin2019computational, constantin2021visual} or perceived complexity \cite{Guo2018complexity}/ 
To facilitate such assessment, several large-scale datasets have been proposed. Examples include AADB \cite{kong2016photo}, AVA \cite{murray2012ava}, AROD \cite{schwarz2018will}, and CADB \cite{zhang2021image} datasets. These datasets provide ground truth annotations for various image quality aspects and serve as resources for training automatic image relevance assessment models. 
However, none of them is made of generated images and provide inputs such as \textit{prompts}.

In the context of the Prompt-Image pairs required for the PPP task, several datasets have emerged. The DiffusionDB dataset 
offers a collection of prompt-image pairs generated using the diffusion models. The Midjourney User Prompts \& Generated Images dataset \footnote{http://bit.ly/kaggle250kMidJourneyDataset} 
provides a diverse set of prompts and corresponding generated images. Additionally, the Dall-E 2 gallery \footnote{https://dalle2.gallery/} 
consists of an online large-scale collection of images generated by the DALL-E 2 model accessible via user interface only. We scrape this former one to build a novel dataset focused on Dall-E 2 generated images. These datasets allow for the evaluation of prompt performance in generative IR systems and serve as valuable resources for training and assessing performance prediction models.
Unfortunately, the text-image datasets lack user relevance feedback. To overcome this drawback, 
our approach involves leveraging pre-trained models on actual images and real human judgments to serve as proxies for generating the missing human relevance scores in our datasets.

\begin{figure}
    \centering \centerline{\includegraphics[width=0.99\columnwidth]{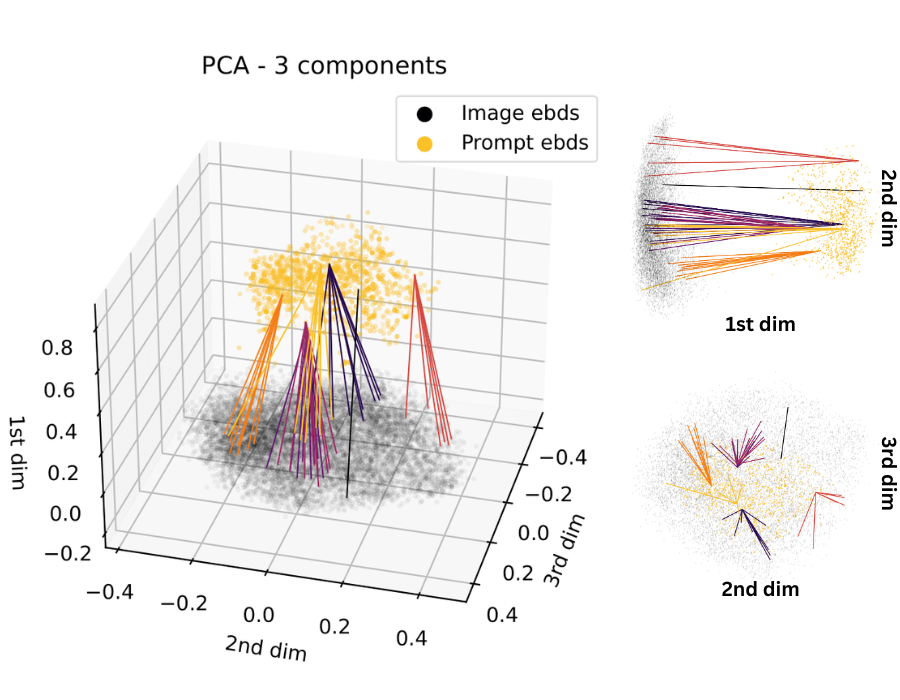}}
    \caption{Principal components from PCA were computed on Clip-ViT-B-32 embeddings of prompts and images (Stable Diffusion). The first component distinctly captures the separation between these two modalities. One prompt can be linked to multiple generated images.}
    \label{fig:pca_-_stable_diffusion_-_vit-b-32}
\end{figure}

\section{Problem Formulation}

In the context of generative IR systems, particularly those generating images from textual prompts, the query equivalent is a prompt that leads to the generation of new content rather than retrieving an existing one. This nuance introduces a change in the QPP task, warranting the introduction of a new task that we refer to as Prompt Performance Prediction (PPP).  
In traditional QPP, the performance of a query is usually estimated based on features of the query and the document corpus, and sometimes top-retrieved documents. However, in our PPP task, the performance must be predicted based on features of the prompt only before generating the images.

Let us formalize the PPP task. Given a generative model $M$, a textual prompt $T_i$, and a set of $J$ images $I=\{I_{i,1}, I_{i,2}, \ldots, I_{i,J}\}$ generated by the model $M$ in response to the prompt $T_i$, we aim to predict the performance of the prompt, denoted as $R(T_i)$. 
We consider that the prompt relevance ground truth can be estimated with the mean of all image relevance scores $R(I_{i,j})$ associated to the prompt $T_i$: 
$ R(T_i) = \frac{1}{J} \sum_{j=1}^{J} R(I_{i, j}) $.

We assume that the true performance $R(T_i)$ is a random variable with mean $f_{\theta}(T_i)$ and variance $\sigma^2$, such that $R(T_i) = f_{\theta}(T_i) + \epsilon$, where $\epsilon \sim \mathcal{N}(0,\sigma^{2})$ is normally-distributed noise. We further assume that the observed performances are samples from this distribution and frame the PPP task as a regression task. 
The goal is to learn a function $f_{\theta}$ parameterized by $\theta$ such that $f_{\theta}(T_i) \approx R(T_i)$ for all prompts in a training set. This function, or performance predictor, is then used to estimate the performance of new prompts.

The parameters $\theta$ of the performance predictor $f_{\theta}$ are learned by maximizing the likelihood of the observed performances given the prompts. Given a dataset of $N$ prompt-performance pairs $(T_i, R(T_i))$, this can be formulated as: 
\begin{equation*}
    \theta^* = \argmax_{\theta} \sum_{i=1}^{N} \log p(R_i | T_i; \theta)
\end{equation*}

This formulation of the PPP task allows for a proactive approach to performance prediction in generative IR systems. By predicting the performance of prompts before images are generated, we can guide the generation and improve the efficiency and effectiveness of the system as a whole.
Such a formulation is indeed compatible with \textit{plug \& play} approaches \cite{Dathathri2020Plug} used in text generation and could be used for prompt reformulation.

\begin{figure}

    \centering 
    \includegraphics[scale=0.55]{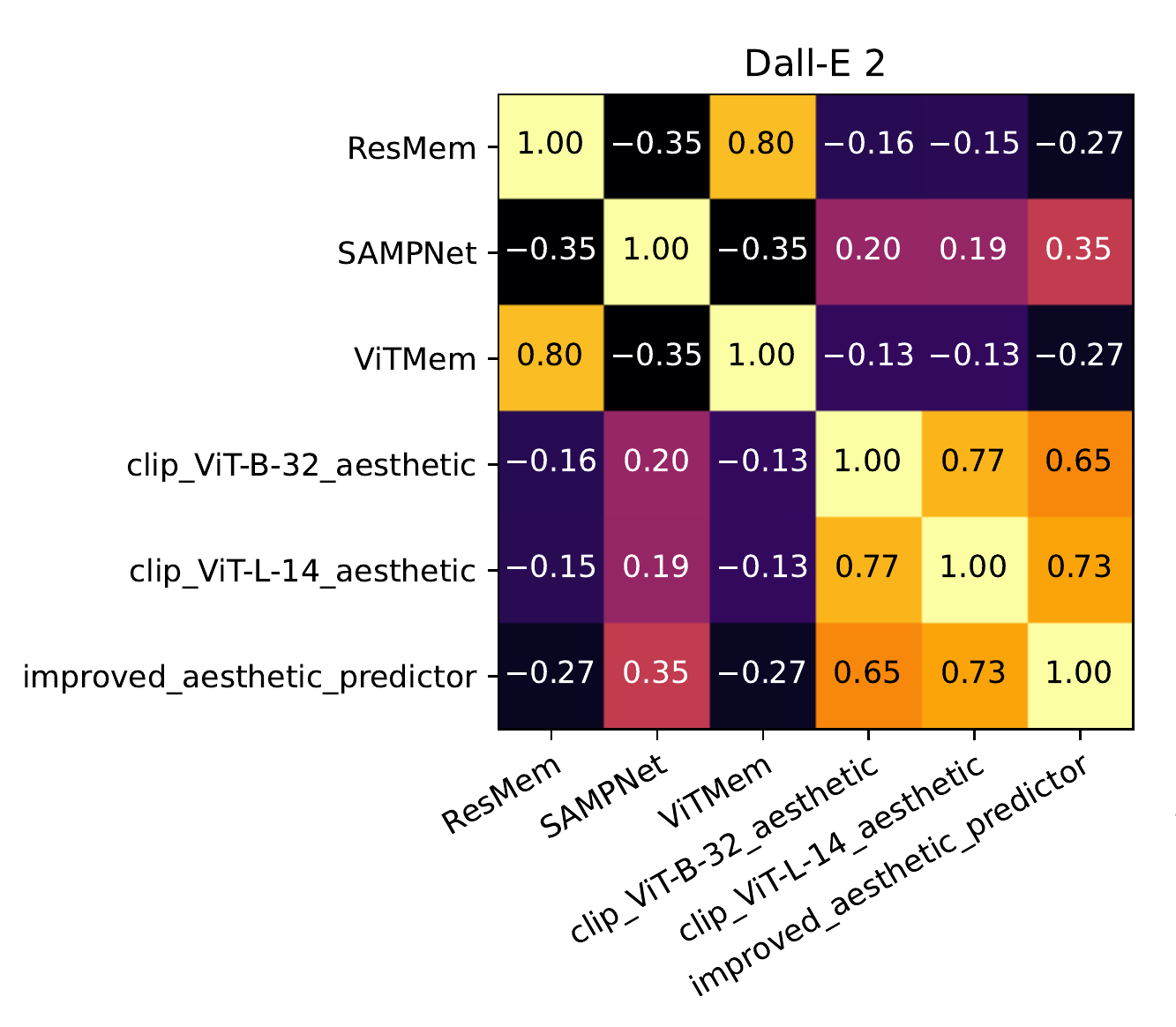}
    \caption{Pearson correlation between image relevance metrics on Dall-E 2: SAMPNet (compositionality), ResMem and ViTMem (memorability), and others (aesthetic). All correlations are statistically significant ($p$-value $< 0.01$)}
    \label{fig:pearson_correlation_between_image_quality_metrics}
\end{figure}

\section{Methodology}

In this section, we present the methodology adopted for our research, which encompasses the creation of datasets with prompt-image-score triplets, the measurement of relevance, and the benchmarking of predictive performance.

\textbf{Images and Prompt Gathering:}
To establish the foundation for our research, we curated three distinct datasets: Midjourney Dataset, Stable Diffusion Dataset, and DALL-E 2 Dataset (name is based on the generator used). Midjourney dataset was crafted from a larger one, incorporating diverse user interactions such as generating variations and upscaling. Stable Diffusion Dataset, on the other hand, represents a subset extracted from the DiffusionDB 
whose the size goes up to 6.5~TB. We sampled this subset from the dataset because of memory and computational limitations. 
Finally, DALL-E 2 Dataset was meticulously obtained through web scraping from an online image database. The Table.~\ref{tab:datasetstats} provides 
the number of prompts and images within each dataset.

\textbf{Ground Truth Relevance Creation:} 
To gauge the relevance of the generated images, we relied on specific criteria, including aesthetic appeal, memorability, and image compositionality. Given the absence of human judgments, we leveraged state-of-the-art pre-trained models to extract relevant scores \cite{zhang2021image, ResMem2021, hagen2023image}  
and Github models \footnote{https://bit.ly/LAION-aesthetic-predictor} 
\footnote{https://bit.ly/improved-aesthetic-predictor} 
based on \cite{hentschel2022clip} findings. 
A total of six distinct pre-trained models were employed, with three dedicated to aesthetic assessment, two to memorability evaluation, and one to compositionality analysis (cf. Table~\ref{tab:linear}). For each neural image grader, by aggregating the scores assigned to images sharing the same prompt, we constructed tuples consisting of prompts and the six corresponding relevance ground truth scores. 
On Figure~\ref{fig:pearson_correlation_between_image_quality_metrics} ones can observe the correlation between the different scores extracted for each dataset. 
This sanity check allow us to confirm previous findings from \cite{Wallraven2015} which report a slight negative correlation between image aesthetics and memorability.

\textbf{Prompts Feature Extractors Benchmarking:}
To evaluate the predictive performance of textual features, we conducted a comprehensive analysis employing eights pre-trained textual feature extractors 
listed in Table.~\ref{tab:linear}. 
We assessed their ability to predict prompt performance through a linear probe evaluation, a well-established protocol in representation learning. By comparing the performance of different pre-trained textual feature extractors, we aimed to identify the most effective approaches in predicting prompt performance accurately and reliably.
Scores reported are Pearson correlation. (All $p$-value $< 0.01$).

\begin{table}[t]
\centering
\caption{Three caption-image datasets used. Each prompt can be used once or several times to produce different images.}
\begin{tabular}{llcc}
\toprule 
                 & Dall-E 2. & Midjourney & Stable Diff  \\
\midrule
        Total prompts      & 27,978                        & 192,080                         & 50,000                                 \\
        Unique prompts     & 17,440                        & 38,349                          & 6,975 \\
\bottomrule
\end{tabular}
\label{tab:datasetstats}
\end{table}

\begin{table*}
\centering
\caption{Pearson correlation scores with linear probe evaluation on top of extracted prompt embedding (All $p$-value $< 0.01$).}  
\resizebox{2.05\columnwidth}{!}{
\begin{tabular}{llcccccccc}
\toprule
    Dataset & Feature Extractors & bart  & gpt2  & bloom560m & sentence-t5-xl  & sentence-t5-base & sentence-t5-large & clip\_ViT-B-32\_prompt  & clip\_ViT-L-14\_prompt \\ 
\midrule
Dall-E 2 & \cite{ResMem2021} ResMem &  0.5771 &  0.5236 & 0.5274 &   0.6158 &  0.6007 &   0.6108 &   \textbf{0.6386} &  \textbf{0.6386} \\
 & \cite{zhang2021image} SAMPNet &  0.4907 &  0.4334 & 0.4366 &   0.5451 &  0.5202 &  0.54 &   0.5684 &   \textbf{0.5746} \\
 & \cite{hagen2023image} ViTMem &  0.6472 &  0.5874 & 0.5958 &   0.6932 &  0.6727 & 0.687 &   0.7136 &   \textbf{0.7166} \\
 & {\small{(3)}} clip\_ViT-B-32\_aesthetic &   0.541 &  0.4682 & 0.4443 & 0.592 &  0.5841 &   0.5991 &   0.6236 &   \textbf{0.6336} \\ 
 & {\small{(3)}}  clip\_ViT-L-14\_aesthetic &  0.5507 &   0.479 & 0.4669 & 0.596 &  0.5903 &   0.6026 &   0.6188 &   \textbf{0.6302} \\ 
 &  {\small{(4)}} improved\_aesthetic\_pre… &  0.5823 &  0.5049 & 0.4828 &   0.6268 &  0.6082 &   0.6294 &   0.6436 &   \textbf{0.6533} \\ 
MidJourney & ResMem &  0.7058 &  0.6712 & 0.6657 &   0.7668 &  0.7488 &   0.7624 &   \textbf{0.8114} &   0.8023 \\
 & SAMPNet &  0.6353 &  0.5863 & 0.5879 &   0.7043 &   0.686 &   0.6967 &   \textbf{0.7726} &   0.7598 \\
 & ViTMem &  0.7271 &  0.6928 & 0.6856 &   0.7878 &  0.7714 &   0.7843 &   \textbf{0.8347} &   0.8268 \\
 & clip\_ViT-B-32\_aesthetic &  0.6013 &  0.5286 & 0.5166 &   0.6665 &  0.6546 &   0.6648 &   \textbf{0.7409} &   0.7313 \\
 & clip\_ViT-L-14\_aesthetic &  0.5978 &  0.5336 &  0.524 & 0.666 &  0.6458 &   0.6611 &   0.7202 &   \textbf{0.7208} \\
 & improved\_aesthetic\_predictor &  0.6584 &  0.5659 & 0.5683 &   0.7097 &  0.6876 & 0.703 &   0.7602 &   \textbf{0.7608} \\
Stable Diff & ResMem &  0.5811 &  0.5248 & 0.5353 &   0.6613 &  0.6378 &   0.6528 &   0.7179 &   \textbf{0.7297} \\
 & SAMPNet &   0.525 &  0.4319 & 0.4416 &   0.6132 &  0.5947 &   0.6092 &   0.6745 &   \textbf{0.6983} \\
 & ViTMem &  0.6249 &  0.5668 & 0.5794 &   0.7043 &  0.6795 &   0.7008 & 0.7670 & \textbf{0.7790} \\
 & clip\_ViT-B-32\_aesthetic &  0.5244 &  0.4273 & 0.4369 &   0.5596 &  0.5538 &   0.5807 &   0.6423 &   \textbf{0.6662} \\
 & clip\_ViT-L-14\_aesthetic & 0.56 &  0.4646 & 0.4727 &   0.5769 &  0.5834 &   0.5969 &   0.6535 &   \textbf{0.6706} \\
 & improved\_aesthetic\_predictor &  0.7215 &   0.636 & 0.6521 &   0.7457 &  0.7444 &   0.7536 &   0.7977 &   \textbf{0.8112} \\
\bottomrule
\end{tabular}
}
\label{tab:linear}
\end{table*}

\sethlcolor{palepurple}

\begin{table}[!ht]
  \centering
  \vspace{-5mm}
  \caption{Scores using the CLIP image embeddings linear aesthetic predictor on prompt embeddings of our datasets. All $p$-values are statistically significative ($< 0.01$) except for the \hl{purple} cell with a $p$-value $=0.5018$.}
         \begin{tabular}{lccc}
            \toprule
            & DALL-E 2 & Midjourney & Stable Diff \\
            \midrule
            CLIP-ViT-B-32 & 0.0347 &
            0.2483 & 
            0.0545 \\ 
            CLIP-ViT-L-14 & \cellcolor{palepurple}-0.0040 &
            0.2173 &
            0.0529 \\
            \bottomrule
        \end{tabular}
  \label{tab:correlation_drop} 
  
  \vspace{-3mm}
\end{table}

\section{Experimental Results}

\textbf{Main Results:} Table~\ref{tab:linear} provides a comprehensive comparison of various textual feature extractors and their performance in the Prompt Performance Prediction (PPP) task, evaluated through linear probe analysis. We present several observations from the experimental results:

\textbf{1)} The Dall-E 2 Dataset poses the greatest difficulty in terms of prediction, exhibiting correlation scores ranging from $0.5746$ to $0.7166$. This can be attributed to the discrepancy in prompt length compared to the other datasets. Notably, the Dall-E 2 model is widely popular among users with low prompt engineering knowledge, as evidenced by the fact that $56.47\%$ of the prompts in Dall-E 2 Dataset contain no modifiers, while in Stable Diff Dataset, they account for only $27.41\%$. 
We believe that the lack of modifiers implies less control and increased randomness, resulting in greater difficulty in forecasting the quality of the produced content.
We confirmed this hypothesis by conducting a Levene's test, comparing the standard deviation of aesthetic scores (using CLIP-L-14 on Dall-E 2 Dataset) between images without modifiers and images with at least one modifier. 
Images with a higher count of modifiers demonstrated superior aesthetic quality and less variance ($p$-value $< 0.01$).

\textbf{2)} CLIP models, as introduced by Radford et al. \cite{radford2021learning}, demonstrate superior performance in the PPP task. This can be attributed to the contrastive pre-training of CLIP on text and image data, which enables the textual component to possess a better "intuition" about the visual meaning of words - an aspect that proves challenging for conventional language models. Additionally, language models are typically trained on natural language corpora, while prompt formulations deviate from traditional grammar by predominantly consisting of a juxtaposition of modifiers. Nevertheless, it is worth noting that the \textit{sentence-t5-xl} model remains highly competitive. 

\textbf{3)}~Assessing compositionality, as estimated by the ground truth using SAMPNet, is found to be the most challenging aspect, with top scores per dataset ranging from $0.5746$ to $0.7726$. In contrast, memorability, estimated by ViTMem, achieves higher top scores per dataset ranging from $0.7166$ to $0.8347$. This discrepancy can be attributed to the strong correlation between memorability and the topic of the image, as established in prior research \cite{hagen2023image}. The topic is easily describable with words, while capturing the geometric and photographic aspects of the image remain more challenging when formulating prompts.

\textbf{Discussion on CLIP Space Discrepancy:} 
Initially, it may appear that pre-trained image aesthetic ground truth predictors are linear models trained on top of CLIP representations. Consequently, one might expect that replicating prompt text scores using textual features from CLIP would be a straightforward task, given its objective of unifying image and text representations within a shared space. However, recent research \cite{goel2022cyclip, liang2022mind} has uncovered a noteworthy discrepancy between the learned representations of images and text in CLIP. 
Contrary to previous assumptions, these representations are not entirely interchangeable and can yield inconsistent predictions in downstream tasks \cite{goel2022cyclip}. This phenomenon, known as \textit{modality gap}, has been  investigated in multi-modal models like CLIP \cite{liang2022mind}. 
The presence of this modality gap significantly impacts the performance on downstream tasks.

In order to gain a comprehensive understanding of the challenges associated with our specific objective, we devised two experiments. 
The first one consist in a visual analysis, as depicted in Figure~\ref{fig:pca_-_stable_diffusion_-_vit-b-32}, showcasing the application of Principal Component Analysis (PCA) on both prompts and images. Through this analysis, we not only validated the previous findings of \cite{goel2022cyclip} regarding the distinct subspaces occupied by prompts and images, but also revealed an intriguing observation: the separation between prompts and images primarily occurs along the first component of the PCA. 

In the second experiment, we sought to predict the aesthetic scores for each dataset by utilizing the aesthetic model as the ground truth extractor, not on image embeddings as previously employed, but rather on the prompt embedding derived from the corresponding clip extractor. The scores presented in Table~\ref{tab:correlation_drop} reveal a severe drop in performances. 
For instance, when utilizing CLIP-B-32, the score for the Midjourney Dataset decreased from $0.7409$ to $0.2483$.

\section{Complementary Experiment}
\label{sec:complementary}

While our first experiment show promising results, they can be undermined by the fact that ground truth scores are generated automatically. 
This is due to the fact that, to our knowledge, no dataset proposes the tuple: \textit{(human prompt, human score)}. 
However, while we explored the tuple: \textit{(human prompt, automatic score)}, to counterbalance we propose a complementary experiment on the setup : \textit{(automatic prompt, human score)} on a different domain: art / paintings.

\begin{figure}
    \centering
    \includegraphics[width=0.48\textwidth]{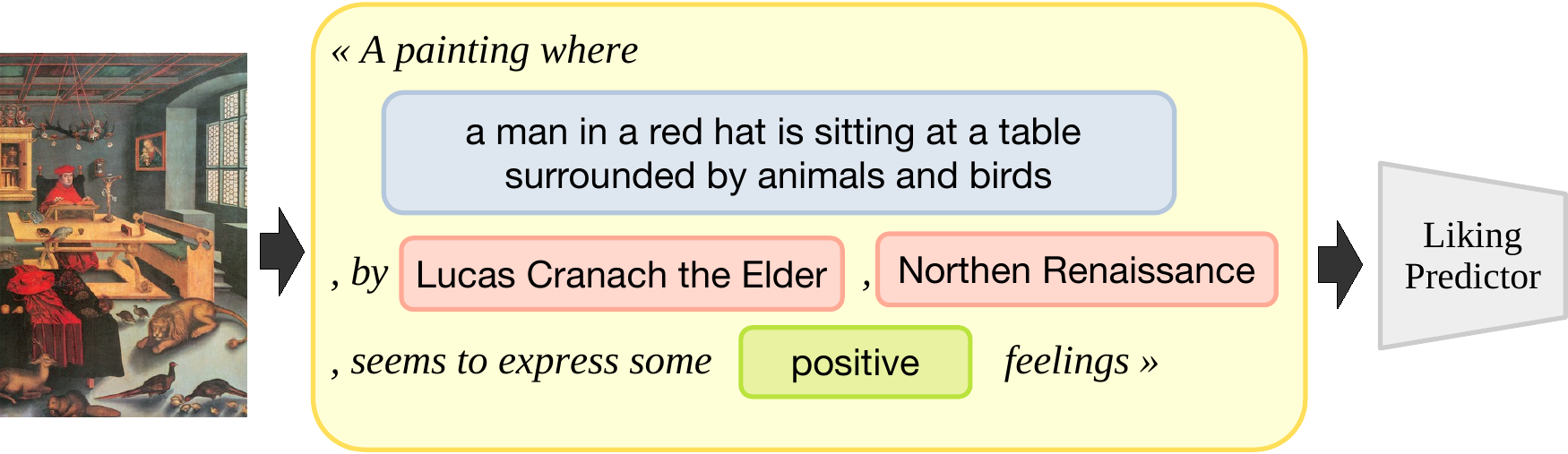}
    \caption{A reconstructed prompt generated from a combination of different information types, including the \textcolor{babyblueeyes}{caption}, \textcolor{candypink}{painter and epoch}, and \textcolor{applegreen}{valence}. The prompt is created based on a general and arbitrary template. 
    When the \textcolor{babyblueeyes}{caption} information is missing, the \textcolor{babyblueeyes}{generated caption} is obtained using the BLIP Image Captioner, 
     which takes the painting image as input. Other information, such as \textcolor{candypink}{painter and epoch}, is extracted from meta-data whenever available, or is inferred from the painting image using other models. 
    }
    \label{fig:pipeline}
\end{figure}

\begin{table*}
\centering
\resizebox{1.95\columnwidth}{!}{
\begin{tabular}{|l||c|c|c|}
\hline
Dataset                   & JenAesthetics \cite{amirshahi2015jenaesthetics}    & (Sidhu, et al.) \cite{sidhu2018prediction} & VAPS-999 \cite{fekete2022vienna}   \\ \hline \hline
Total pictures $|$ Missing            & 1628  $|$  65   (3.99\%)     & 480   $|$  6 (1.25\%)                                               & 999   $|$  3  (0.30\%)                                             \\ \hline
Total participants        & 134                                                                 & 598                                                         & 120 (60 M, 60 F)                                    \\ \hline
Diversity of participants & \makecell{15 countries \\ (109 from Germany)}                                     &          \makecell{University of Calgary \\ undergraduates}                                                   & \makecell{ University of Vienna students \\ 60 M, 60 F} \\ \hline 
Styles covered            &  Renaissance $\xrightarrow{}$ Expressionism                                   & \makecell{Half abstract \\ half representationnal}                       & 1434 $-$ 2012                                         \\ \hline
Community                 & Computer Vision                                                     & Psychology                                                  & Psychology \& Art                                   \\ \hline
Appreciation features     & \makecell{Aesthetic, Beauty, \\ Liking Color / Composition / Content} & \makecell{Meaningfulness, Complexity \\ Emotion, Beauty, Liking}                                       & Cognitive, Affective, Liking                      \\
\hline
\end{tabular}}
\caption{The datasets of paintings used in the complementary experiments detailed in Section \ref{sec:complementary}, each associated with liking ratings and appreciation. These datasets cover a wide range of styles and epochs with diverse communities of users for rating.}
    \label{tab:paintings_datasets}
\end{table*}

\subsection{Complementary Experiments Setup}

This experiment aims to predict the appreciation and human rating of paintings solely using captions, rather than relying on image-based features. To accomplish this, and as our goal is only to prove the usefulness of language as a way to understand appreciation, we employ a straightforward training approach that utilizes text descriptions of each painting, coupled with their corresponding rating of appreciation, as our input data. Our model consists of a text feature extractor paired with a linear regressor, which we fit to predict the appreciation ratings of each painting using their description. However, we often face the issue of a lack of descriptions for some paintings. To overcome this, we utilize pre-trained generative models to generate captions of painting images.

In order to further diversify and complement the information present in the description, we extract useful meta-data from each painting. This meta-data can include the name of the artist, the genre of art, and valence, which represents the emotional tone conveyed by the art (whether it has a positive or negative feeling). To generalize our approach to text-to-image generative models, we merge all of this information into prompts commonly used in these models, following the general template described in Figure \ref{fig:pipeline}. This allows us to create a comprehensive prompt that can be used to predict appreciation and human rating for any given painting.

To this end, we leverage a CLIP ViT-L-14 text encoder in conjunction with a linear regressor, as previously demonstrated in \cite{hentschel2022clip}. However, no existing datasets feature human ratings alongside image captions for each painting. To address this, we focus on appreciation rating datasets and utilize the BLIP Image Captioner\footnote{\url{https://github.com/salesforce/BLIP}} to generate text descriptions (i.e., captions) from painting images. We finetune the BLIP Captioner on the Objective Language for Art (OLA) dataset caption-image pairs, which contain 5,000 unique pairs. 
In this way, we augment existing datasets, namely JenAesthetics, VAPS-999, and Sidhu, which include both paintings and the corresponding set of human liking ratings, with text descriptions for each painting. We use a separate linear regressor to predict each appreciation feature. 

In our experiments, we also extract additional meta-data from the painting images, such as the valence (emotional feeling), the artist and the epoch of painting. 
If this information is not provided, we utilize a linear classifier 
trained on the WikiArt dataset\footnote{\url{https://www.wikiart.org}} to extract information about the artist and the epoch of the painting. We then combine this information with the painting captions to form prompts as in Figure \ref{fig:pipeline}. 

Paintings appreciation is a different task from aesthetic ratings and poorly correlated. 
This makes our complementary analysis even more interesting by bringing a novel task/metric and different domain (paintings/art). 
We employ the LAION aesthetic predictor as a baseline. The LAION aesthetic predictor, which concatenates the CLIP ViT-L-14 feature extractor with a linear regression layer, is trained on the Simularcra Aesthetic Caption (SAC)\footnote{\url{https://github.com/JD-P/simulacra-aesthetic-captions}} and the AVA photography dataset \cite{murray2012ava} to predict the aesthetic of an image (which has been found to be different from his appreciation). 
To ensure fair comparison, we employ the CLIP ViT-L-14 image encoder, which is coupled with the same text encoder used in our approach, and train it using a linear regressor 
on the three datasets presented in Table \ref{tab:paintings_datasets} to predict human liking ratings. 

\begin{table}[]
    \centering
    \resizebox{0.99\columnwidth}{!}{
    \begin{tabular}{|l|c|c|c|}
        \hline
        Method & VAPS-999    &  Sidhu & JenAesthetics \\ 
        \hline \hline
        LAION Aesthetic Predictor & \textbf{0.2232}  & \textbf{0.2938} & \textbf{0.1635} \\ \hline 
        \hline
        Text-based : \textcolor{applegreen}{Valence} & 0.4025 & 0.5980 & $-$ \\ \hline
        Text-based : \textcolor{babyblueeyes}{Description} & 0.4401  & 0.3689 & 0.5452 \\ \hline
        Text-based : \textcolor{candypink}{Painter + epoch} & 0.5471  & 0.4185 & 0.4826 \\ \hline
        \makecell[l]{Text-based : \textcolor{babyblueeyes}{Description} \\ + \textcolor{candypink}{painter + epoch}} & 0.5989  & 0.4955 & \textbf{0.5805} \\ \hline
        \makecell[l]{Text-based : \textcolor{babyblueeyes}{Description} \\ + \textcolor{candypink}{painter + epoch}  +  \textcolor{applegreen}{valence}} & \textbf{0.6581} & \textbf{0.6204} & $-$ \\ \hline\hline
        Image-based  features  & \textbf{0.7269} & \textbf{0.7939} & \textbf{0.7366} \\ \hline
    \end{tabular}}
    \caption{Pearson correlation results between the appreciation ground truth and appreciation predictions of the paintings of each of the three datasets \cite{amirshahi2015jenaesthetics,fekete2022vienna,sidhu2018prediction} of our study (all $p$-value $< 0.01$), made using the Laion Aesthetic Predictor pretrained on photography and natural images (top),  our proposed text-based approach with varying degrees of information for prediction using text-based features of paintings (middle), and an image prediction model with image-based features of the paintings (bottom). The text-based and image-based approach use the same CLIP ViT-L-14 encoder \cite{radford2021learning} coupled with a linear layer for regression.}
    \label{tab:scores_vit14}
    \vspace{-3pt}
\end{table}

\subsection{Complementary Results}

Table \ref{tab:scores_vit14} presents all pearson correlation results between our predictions and the ground truth, which are all statistically significant with a $p$-value $< 0.01$. Our findings indicate that, although using text does not surpass the use of image features, the difference in performance is relatively small, demonstrating the potential of using prompts alone to predict the appreciation of paintings. Moreover, prompt engineering shows a significant impact on the quality of predictions. The use of multiple types of information, when available, results in a significant improvement in the quality of the results. Our findings also reveal that while the information regarding the painter and epoch plays a crucial role in determining the appreciation of a painting, additional information that is specific to the painting itself, such as its description, is required to achieve a significant improvement in the quality of the results. We further observe that the effect of each information source in the prompt differs across datasets since each dataset exhibits different art styles that may be difficult to describe through a caption, but can be captured well with the painting's valence and emotion. Interestingly, using all available information together in a prompt yields relatively close results across all three datasets, despite the high variance in results for individual information sources. This suggests that, even in different painting styles, all information sources can complement each other to complete the information missing in each individual source. Finally, we show that using the LAION predictor, 
which is pretrained on photography and natural images, yields significantly lower results on all three datasets compared to using either images or text descriptions of paintings, indicating that the features used to predict appreciation in photography and natural images hold significant semantic differences from painting features, and that they are challenging to reconcile.

Upon conducting a thorough investigation of the composition of the datasets utilized in our complementary analysis, we observed notable variations in the types and styles of the artworks featured in each. For instance, approximately 50\% of the paintings in the Sidhu dataset are of the abstract style, while the JenAesthetics dataset is exclusively composed of figurative paintings. This disparity may provide an initial explanation for the significant impact of the description on the predictions of appreciation for JenAesthetics, as well as the challenges involved in using descriptions alone to predict the appreciation of abstract paintings. Moreover, it is plausible that the information pertaining to the painter and epoch of the painting might inherently provide a better explanation for the appreciation of abstract paintings.

Furthermore, appreciation levels can differ significantly among different subgroups of people, thus resulting in widely varying appreciations. For example, the VAPS-999 study and dataset \cite{fekete2022vienna} divided all its participants into six subgroups prior to giving their appreciations, with the correlations between the appreciations of each subgroup ranging from 0.607 to 0.821. Given that our predictions' correlations with the ground truth fall within this range, it is reasonable to infer that the text-based approach attains a performance similar to human-level appreciation using painting images on the VAPS-999 dataset.

This complementary analysis demonstrates the feasibility of predicting image appreciation from its textual description. 

\section{Conclusion and Future Work}

In this study, we extended the traditional Query Performance Prediction (QPP) task to image generation by proposing the new Prompt Performance Prediction (PPP) task. 
Our proposed framework assesses the effectiveness of textual prompts in generating images based on aesthetics, memorability, and compositionality. Surprisingly, even simple linear predictors fine-tuned on pretrained textual encoders proved efficient. 
We found our approach to be efficient on both complementary setups : generated prompt, real scores and real prompt automatic scores on two modalities (images and paintings). 
This approach can empower users to evaluate the potential effectiveness of their prompts before investing time and resources in content creation. PPP could also be integrated into frameworks like \textit{plug \& play} approaches for automatic prompt reformulation and for identifying biases by analyzing how prompt words influence scores. Lastly, we recommend creating new datasets with real human judgments to gain a deeper understanding, as automated models may miss certain nuances.


\bibliographystyle{IEEEbib}
\bibliography{strings,refs}

\end{document}